\documentclass[%
 aip,
 jmp,%
 amsmath,amssymb,
reprint,%
]{revtex4-1}

\usepackage{graphicx}
\usepackage{dcolumn}
\usepackage{bm}
\usepackage{float}
\usepackage{color}
\usepackage[draft]{changes} 
\usepackage[normalem]{ulem}
\definechangesauthor[name={yi}, color=blue]{Yi}
\definechangesauthor[name={Maria}, color=red]{Maria}

\usepackage{array}
\newcolumntype{P}[1]{>{\centering\arraybackslash}p{#1}}
\newcommand{\etal}{\textit{et al.}}

\usepackage{graphicx}
\usepackage{sidecap}
\usepackage{tabularx}

\begin{document}
\title{Anharmonic stabilization and lattice heat transport in rocksalt $\beta$-GeTe}

\author{Yi Xia} 
\email{yxia@anl.gov}
\affiliation{Center for Nanoscale Materials, Argonne National Laboratory, Argonne, Illinois 60439, USA}
\author{Maria K. Y. Chan}
\email{mchan@anl.gov}
\affiliation{Center for Nanoscale Materials, Argonne National Laboratory, Argonne, Illinois 60439, USA}

\date{\today}

\begin{abstract}
Peierls-Boltzmann transport equation, coupled with third-order anharmonic lattice dynamics calculations, has been widely used to model lattice thermal conductivity ($\kappa_{l}$) in bulk crystals. However, its application to materials with structural phase transition at relatively high temperature is fundamentally challenged by the presence of lattice instabilities (imaginary phonon modes). Additionally, its accuracy suffers from the absence of higher-than-third-order phonon scattering processes, which are important near/above the Debye temperature. In this letter, we present an effective scheme that combines temperature-induced anharmonic phonon renormalization and four-phonon scattering to resolve these two theoretical challenges. We apply this scheme to investigate the lattice dynamics and thermal transport properties of GeTe, which undergoes a second-order ferroelectric phase transition from rhombohedral $\alpha$-GeTe to rocksalt $\beta$-GeTe at about 700~K. Our results on the high-temperature phase $\beta$-GeTe at 800~K confirm the stabilization of $\beta$-GeTe by temperature effects. We find that considering only three-phonon scattering leads to significantly overestimated $\kappa_{l}$ of 3.8~W/mK at 800~K, whereas including four-phonon scattering reduces $\kappa_{l}$ to 1.7~W/mK, a value comparable with experiments. To explore the possibility to further suppress $\kappa_{l}$, we show that alloying $\beta$-GeTe with heavy cations such as Pb and Bi can effectively reduce $\kappa_{l}$ to about 1.0~W/mK, whereas sample size needs to be around 10nm through nanostructuring to achieve a comparable reduction of $\kappa_{l}$.
\end{abstract}

\maketitle



Thermoelectric materials hold promise to alleviate the energy crisis and environmental impacts of energy use, owning to their ability to convert waste heat into electricity. Materials with strong intrinsic lattice anharmonicity has drawn vast attention because of low lattice thermal conductivity ($\kappa_{l}$), which is key for enhancing thermoelectric conversion efficiency. Recently, high thermoelectric performance has been reported in a series of materials with structural phase transition such as SnSe,\cite{Zhao:2014aa} GeTe,\cite{WuDi2014} and Cu$_2$Se,\cite{Liu:2012aa} which is primarily ascribed to the ultralow $\kappa_{l}$ at the high temperature end, suggesting a relationship between lattice instability (phase transition) and strong anharmonicity.

GeTe undergoes a second-order ferroelectric phase transition from rhombohedral $\alpha$-GeTe (space group $R3m$) to rocksalt $\beta$-GeTe (space group $Fm\bar3m$) at the critical temperature ($T_{c}$) of $650\pm100$ K.\cite{STEIGMEIER19701275,Chattopadhyay1987} Recently, GeTe-based thermoelectric materials have been reported to exhibit high thermoelectric performance,\cite{Levin2013,WuDi2014,Madar2016,Suresh2016,SureshJMCC2016,Li:2017aa,LiJuan2017,Suresh2017,MinHong2018,Li:2018aa} showing great potential as a replacement for PbTe-based materials, which contain highly-toxic Pb that limits their practical applications. A very high $zT$ of 2.3 has been reported in Sb and In co-doped $\beta$-GeTe owning to the superior electronic properties and low $\kappa_{l}$.\cite{MinHong2018} It is suggested that the phase transition from the low-symmetry rhombohedral $\alpha$-GeTe to the high-symmetry rocksalt $\beta$-GeTe plays a crucial role in enhancing thermoelectric efficiency through converging band pockets and suppressing $\kappa_{l}$.\cite{WuDi2014,Suresh2017,MinHong2018} Therefore, it is fundamentally important to understand the lattice dynamics and thermal transport in $\beta$-GeTe, insight into which may enable further reduction of $\kappa_{l}$ and enhancement of $zT$. However, existing theoretical studies have thus far only focused on $\alpha$-GeTe.\cite{Davide2017,Bosoni2017} The comprehensive understanding of $\beta$-GeTe is hindered by theoretical challenges originating from lattice instabilities (imaginary phonon modes) and strong higher-than-3rd-order anharmonicity inherent in $\beta$-GeTe. Motived by this point, we perform first-principles-based calculations of lattice dynamics and thermal transport properties of $\beta$-GeTe at 800 K, near the temperature of optimal experimental thermoelectric performance.\cite{MinHong2018} To resolve the imaginary phonon modes, we present an effective scheme to account for temperature effect by explicitly including anharmonicity up to 4th order, with a particular focus on temperature-induced anharmonic phonon renormalization (PRN) and four-phonon scattering.


\begin{figure*}[htp]
	\includegraphics[width = 0.85\linewidth]{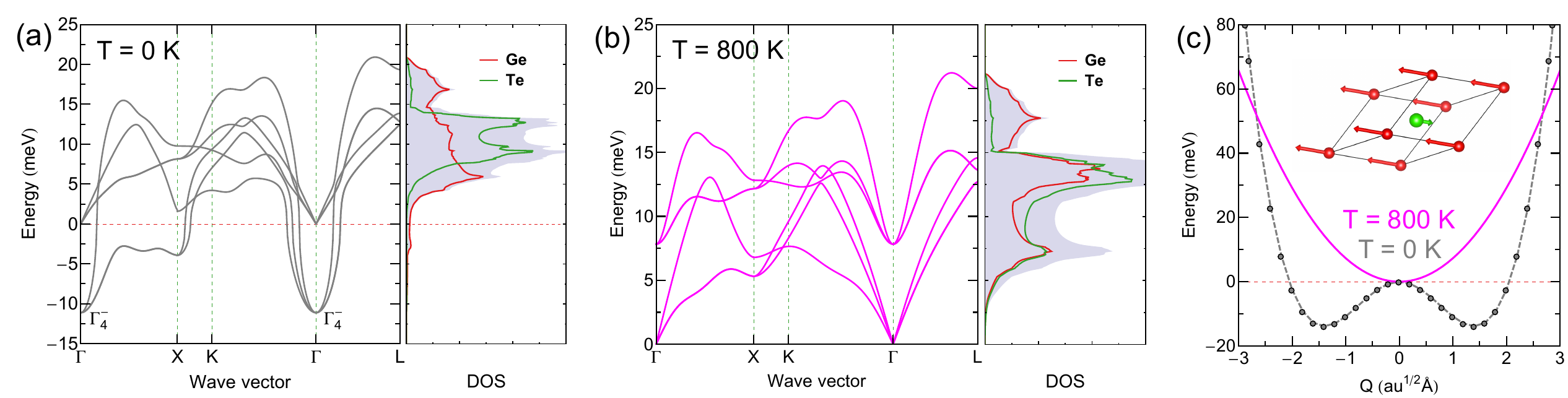}
	\caption{ Phonon dispersions and atom-projected densities of states for rocksalt $\beta$-GeTe computed at 0 K (a) and 800 K (b), respectively. Imaginary phonon frequencies are denoted by negative values. The triply-degenerate optical modes at the $\Gamma$ point are denoted as $\Gamma_{4}^{-}$ following Ref.~[\onlinecite{Wdowik2014}]. (c) Potential energy surface associated with the frozen $\Gamma_{4}^{-}$ mode as a function of vibrational magnitude computed using DFT (gray disks). The gray dashed lines depict a polynomial fit up to $6th$-order. The solid magenta lines correspond to the effective harmonic potential energy surface of the renormalized $\Gamma_{4}^{-}$ mode at 800 K. The inset shows the atomic displacements associated with one of the three eigenvectors of $\Gamma_{4}^{-}$ modes.
	} 
	\label{fig:Dispersion}
\end{figure*}

In order to determine the phonon dispersion in $\beta$-GeTe, we perform density-functional-theory (DFT)\cite{dft} calculations with the projector-augmented wave (PAW) method,\cite{paw} using the atom potentials supplied with the $Ab\ initio$ Simulation Package (VASP).\cite{Vasp1, Vasp2, Vasp3, Vasp4,vaspPAW} Particularly, we used Perdew-Burke-Ernzerhof (PBE)\cite{pbe,Perdew2008} functional, which is a generalized-gradient approximation (GGA) to the exchange-correlation functional.  Supercell (4$\times$4$\times$4) structures were constructed and compressive sensing lattice dynamics (CSLD)\cite{csld,2018arXiv180508904Z,2018arXiv180508903Z,Yi2018PbTe} was utilized to extract both harmonic and anharmonic interatomic force constants (IFCs). Fig.\ \ref{fig:Dispersion}(a) shows the phonon dispersion relation of $\beta$-GeTe at 0~K, where strong phonon instabilities and imaginary phonon frequencies are found near the $\Gamma$ point and along the $\Gamma$-X path in the Brillouin zone. The triply-degenerate zone-center optical modes ($\Gamma_{4}^{-}$) with energy of about $11i$ meV display the strongest instabilities, and the doubly-degenerate modes at the X point have a lower energy of $4.0i$ meV. The mass-reduced polarization vector associated with one of the triply-degenerate $\Gamma_{4}^{-}$ modes, shown in Fig.\ \ref{fig:Dispersion}(c), reveals large magnitude of atomic displacements associated with a stretch of the Ge-Te bond. The corresponding potential energy surface (PES) exhibits a double-well shape with the rocksalt structure at the dynamically-unstable saddle point and minimum energies 14.4 meV below the saddle point. Our results are consistent with a previous first-principles study by Wdowik \textit{et al},\cite{Wdowik2014} which identified similar phonon dispersion curve and double-well PES.

Physically, these imaginary phonon frequencies are at odds with the experimentally-observed stability of $\beta$-GeTe above $T_{c}$, which reveals that harmonic approximation is no longer valid and anharmonic effects are pronounced with sufficiently large displacive amplitudes. It is evident from the double-well PES that high-order anharmonicity with at least a quartic term is required to stabilize $\beta$-GeTe above $T_{c}$. To obtain properly-renormalized phonon frequencies at finite temperature, we adopt a phonon renormalization (PRN) scheme based on iteratively refining temperature-dependent effective harmonic IFCs. This scheme is similar to the temperature dependent effective potential (TDEP)\cite{TDEP} method in the sense that both methods extract effective harmonic IFCs that best fit the forces experienced by atoms at finite temperature. Our PRN scheme consists of two essential steps: (1) generating a thermal distribution of atomic displacements according to the given temperature and calculating the residual forces experienced by atoms, and (2) fitting effective harmonic IFCs by minimizing the corresponding force prediction error. Note that in order to obtain properly-renormalized phonon frequencies, only harmonic IFCs are fitted so that higher-order terms are effectively renormalized into the harmonic IFCs. In practice, we start with the harmonic IFCs computed at 0~K to obtain thermalized structures, and then iterate the above two steps by monitoring the convergence of the renormalized phonon frequencies and vibrational free energy. Different from the TDEP method,\cite{TDEP} however, we employ IFCs up to 6th order extracted using CSLD\cite{csld,2018arXiv180508904Z} to efficiently predict forces for thermalized supercell structures, and instead of performing time-consuming first-principles molecular dynamics simulations, we generate temperature-dependent atomic displacements according to a quantum covariance matrix $\Sigma_{a\alpha, b\beta}$ \cite{Errea2014,Roekeghem2016} of atomic displacements,
	\begin{equation}\label{eq:covar}
	\Sigma_{a\alpha, b\beta} = \frac{\hbar}{2\sqrt{m_{a}m_{b}}} \sum_{\lambda} \frac{\left(1+2n^{0}_{\lambda}\right) }{\omega_{\lambda}} e^{\lambda}_{a\alpha} e^{\ast\lambda}_{b\beta},
	\end{equation}
where $m$, $\omega$, $n^{0}$ and $e$ are atomic mass and phonon frequency, population (at the given temperature based on Bose-Einstein distribution), and eigenvector, respectively. The variables $a$/$b$, $\alpha$/$\beta$ and $\lambda$  are indices for atom, cartesian coordinate, and phonon mode, respectively. Numerical details of IFCs fitting, iterative phonon renormalization and a benchmarking study of SrTiO$_{3}$ are given in the supplementary material (SM).

Upon performing the phonon renormalization procedure at 800~K, we find that the energy of the $\Gamma_{4}^{-}$ modes renormalized from 11$i$ meV to 7.8 meV as shown in Fig.\ \ref{fig:Dispersion}(b), demonstrating that $\beta$-GeTe is indeed dynamically stable above $T_{c}$. A simplified interpretation of this anharmonic stabilization in the frozen phonon picture is that the high-temperature $\beta$-GeTe is an effectively averaged structure with atoms moving frequently across nearby local energy minima, and therefore, the effective harmonic PES associated with $\Gamma_{4}^{-}$ modes at 800 K is positive definite, as shown in Fig.\ \ref{fig:Dispersion}(c). Compared to the phonon dispersion at 0 K, three optical branches and acoustic modes along $\Gamma$-X and $\Gamma$-K paths are significantly renormalized. With temperature increased from 0 K to 800 K, we find that the phonon density of states (DOS), as shown in Fig.\ \ref{fig:Dispersion}(a) and (b), exhibits several noticeable changes: (1) the low-lying Ge peak at 5.9 meV increases to 7.3 meV and a new Ge peak appears near 13.3 meV, and (2) the low-lying Te peak at 9.1 meV is significantly softened to join the low-lying Ge peak, while high-lying Te peak at 12.8 meV is only slightly hardened. Therefore, we expect significant change in computed $\kappa_{l}$ with PRN, despite the fact the DOS associated with unstable phonons is confined to a small reciprocal-space volume.\cite{MinHong2018}

\begin{figure*}[htp]
	\includegraphics[width = 0.8\linewidth]{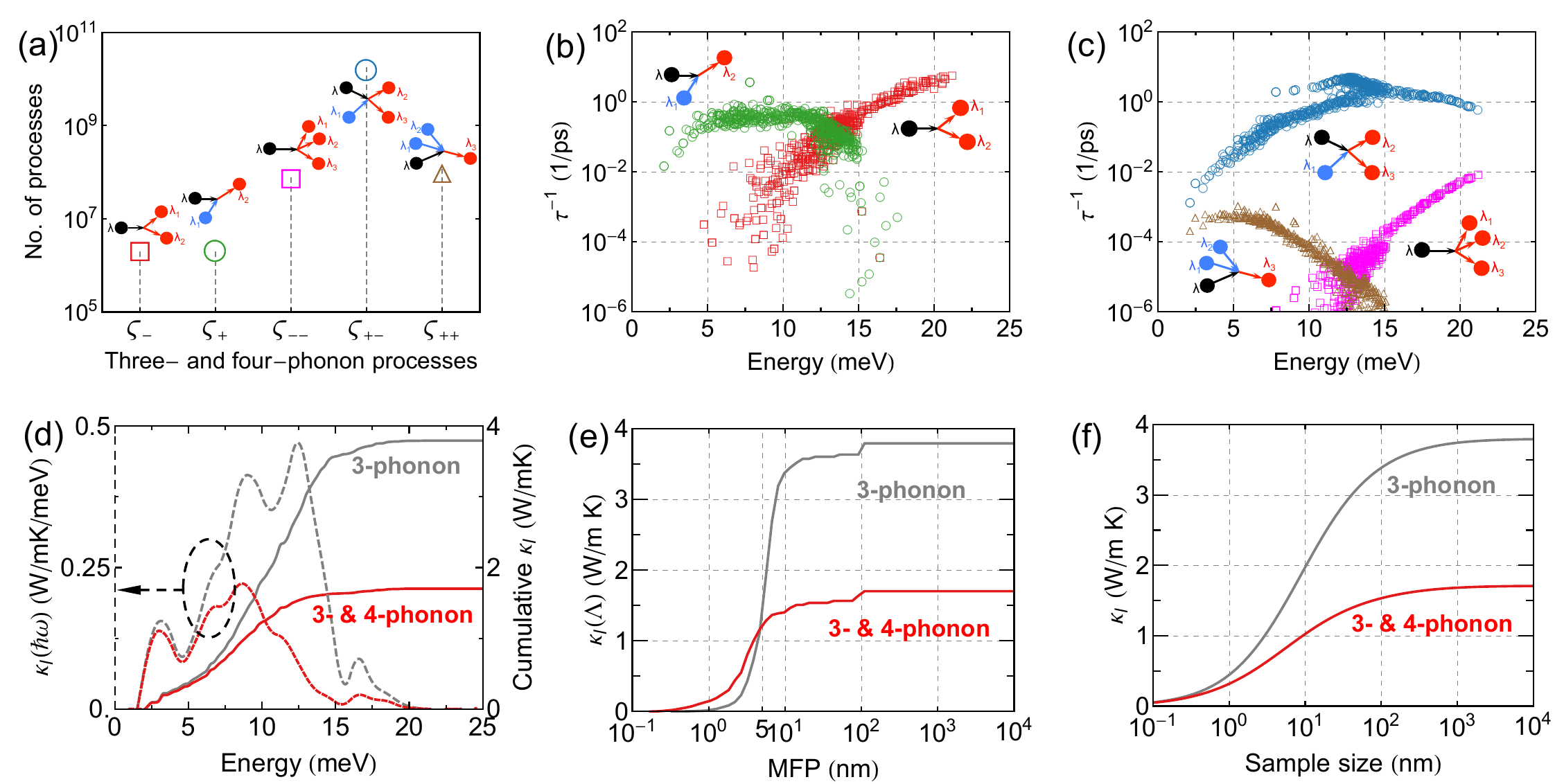}
	\caption{ 
	(a) Number of allowed three- and four-phonon processes that conserve crystal momentum and energy using renormalized phonon dispersion at 800~K. The insets show various types of scattering processes. (b) Mode-dependent three-phonon scattering rates at 800 K, wherein splitting ($\zeta_{-}$) and combination ($\zeta_{+}$) processes are denoted as red squares and green circles, respectively. (c) Mode-dependent four-phonon scattering rates at 800~K, wherein the splitting ($\zeta_{--}$), redistribution($\zeta_{+-}$), and combination ($\zeta_{++}$) processes are denoted respectively as magenta squares, blue circles and brown triangles. (d) Energy cumulative $\kappa_{l}$ (solid lines) and thermal conductivity spectra $\kappa_{l}(\hbar\omega)$ (dashed lines) at 800~K. (e) Mean free path (MFP) cumulative $\kappa_{l}(\Lambda)$ (solid lines) as defined in Eq.(\ref{eq:kappamfp}) at 800~K. (f) $\kappa_{l}$ including phonon-boundary scattering as a function of sample size at 800 K.
	}
	\label{fig:Rates}
\end{figure*}

With the stabilized structure and corresponding phonon properties, it is possible to compute the lattice thermal conductivity. Within the framework of anharmonic lattice dynamics (ALD) and Peierls-Boltzmann transport equation (PBTE)\cite{peierls1996quantum,srivastava1990physics,ziman} under the relaxation time approximation, the lattice thermal conductivity tensor ($\kappa_{l}^{\alpha\beta}$) is expressed as a sum over contributions from phonon modes in the first Brillouin zone,
\begin{equation}\label{eq:kappa}
\kappa_{l}^{\alpha\beta}= \frac{1}{NV}\sum_{\lambda} C_{\lambda} v_{\lambda}^{\alpha} v_{\lambda}^{\beta} 	\tau_{\lambda},
\end{equation}
where $C_{\lambda}$, $v_{\lambda}^{\alpha}$ and $\tau_{\lambda}$ are heat capacity, group velocity and lifetime of phonon mode $\lambda$, respectively. $N$ and $V$ are respectively the number of sampled phonon wave vectors and the volume of the primitive cell. To calculate $\kappa_{l}$, $C_{\lambda}$ and $v_{\lambda}^{\alpha}$ is extracted from phonon dispersion, and $\tau_{\lambda}$ is typically computed by considering anharmonic three-phonon scattering (lowest order in perturbation theory). Recently, Feng~\etal\cite{Tianli2016,Tianli2017} and Thomas~\etal\cite{Thomas2017} have demonstrated that even for diamond, silicon, and germanium, which are weakly anharmonic compounds, scattering rates from four-phonon processes are comparable to three-phonon scattering rates near the Debye temperature. Considering the strong quartic anharmonicity associated with $\beta$-GeTe reflected in the double-well PES, significant four-phonon scattering is expected at 800 K. Therefore, we estimate mode-dependent lifetimes by taking into account both three- and four-phonon scattering processes, as well as isotopic scattering, using Matthiessen's rule. Specifically, we include splitting ($\zeta_{-}:  \lambda \rightarrow \lambda_1 + \lambda_2 $) and combination processes ($ \zeta_{+}: \lambda + \lambda_1 \rightarrow \lambda_2 $) for three-phonon scattering. And for four-phonon scattering, we consider splitting ($ \zeta_{--}: \lambda \rightarrow \lambda_1 + \lambda_2+ \lambda_3 $), redistribution ($ \zeta_{+-}: \lambda + \lambda_1 \rightarrow \lambda_2+ \lambda_3 $), and combination ($ \zeta_{++}: \lambda + \lambda_1 + \lambda_2 \rightarrow  \lambda_3 $) processes [see Fig.\ \ref{fig:Rates}(a)].\cite{Tianli2016} We employ an approximated solution for PBTE, namely, the single mode relaxation time approximation (SMRTA), which could lead to potential underestimation of $\kappa_{l}$ (compared to that obtained by full iterative solution of the PBTE) if phonon scattering is dominated by $normal$ processes. However, we find that the underestimation due to SMRTA is negligible when only three-phonon scattering is considered for $\beta$-GeTe at 800~K (see Fig.~S5 in the SM). Due to the much heavier computational burden, we are unable to iteratively solve the PBTE\cite{OMINI1995101,omini2} including four-phonon scattering processes.

It is noteworthy that a recent study by Ju \etal\cite{Shenghong2018} reveals significant long-range effect on the IFCs in rocksalt PbTe due to resonant bonding,\cite{Lee:2014aa} which leads to a strong dependence of phonon dispersion and $\kappa_{l}$ on the number of neighbor shells and number of atoms in supercell. Since $\beta$-GeTe adopts the same structure and orbital configurations as PbTe, we perform additional convergence test using larger supercell (up to 250 atoms) and longer cutoff distance (up to 7th neighbor) for 3rd-order anharmonic interactions. We find that $4\times4\times4$ supercell and including up to the 5th neighbor cubic anharmonicity lead to reasonably good convergence of phonon dispersion and $\kappa_{l}$ when only three-phonon scattering is considered. Due to the combinatorial explosion in the number of parameters associated with the 4th-order IFCs and large computational cost in evaluating four-phonon scattering,\cite{Tianli2016,pbte2018} the 4th-order anharmonicity is limited  to the 2nd nearest neighbor. Detailed formulations of three- and four-phonon scattering rates and convergence tests of $\kappa_{l}$ are included in the SM.

Our results show that four-phonon processes are more numerous than three-phonon ones, and also contribute significantly to the total scatterings. Fig.\ \ref{fig:Rates}(a) displays the number of captured phonon processes that conserve both crystal momentum and energy. With a 14$\times$14$\times$14 mesh of phonon wave vectors, the number of splitting and combination three-phonon processes are on the order of 10$^6$. In contrast, the number of four-phonon processes is significantly larger, with the dominating redistribution processes on the order of 10$^{11}$. Fig.\ \ref{fig:Rates}(b) and (c) show the corresponding mode-dependent phonon scattering rates. In the three-phonon processes, phonon scattering rates of low-energy/high-energy modes are mainly from combination/splitting processes, consistent with the energy conservation constraints.  With respect to the four-phonon processes, the total scattering rates have predominant contributions from redistribution processes due to the large number of scattering events. The comparison between three- and four-phonon scattering rates reveals that, contrary to expectations, for $\beta$-GeTe at 800~K, four-phonon scattering is comparable in strength to three-phonon scattering over the entire energy range. Particularly, four-phonon scattering is even stronger than three-phonon scattering for phonon modes with energies ranging from about 10 meV to 15 meV.

Correspondingly, inclusion of four-phonon processes changes the value of $\kappa_{l}$  significantly. Fig.\ \ref{fig:Rates}(d) shows the computed $\kappa_{l}$ and its spectra $\kappa_{l}(\hbar\omega)$. Including only three-phonon scattering leads to $\kappa_{l}$ of 3.8 W/mK at 800 K, while further including four-phonon scattering significantly reduces $\kappa_{l}$ to 1.7 W/mK, a reduction of nearly 55\%. The calculated $\kappa_{l}(\hbar\omega)$ indicates that contributions to total $\kappa_{l}$ from phonon modes with energies larger than 6.5 meV are significantly suppressed, which is consistent with the observed large four-phonon scattering rates. We find that the values of $\kappa_{l}$ predicted considering both three- and four-phonon scattering rates achieves reasonably good agreement with various experimental reports, \cite{MinHong2018,LiJuan2017,Suresh2017,WuDi2014} as detailed in Table\ \ref{table:KappaGeTe}. The remaining discrepancy may be attributable to the scattering from experimentally identified Ge vacancies.\cite{Levin2013,Li:2017aa,Suresh2017,LiJuan2017} We further estimate the effect of Ge vacancies on $\kappa_{l}$ by treating it perturbatively as an isotopic effect according to Ref.~[\onlinecite{Ratsifaritana1987},\onlinecite{Davide2017}]. In this estimate, Ge vacancies at 2.1 atomic~\%, corresponding to a hole concentration of 7.8$\times10^{20}$~cm$^{-3}$ in Ref.~[\onlinecite{LiJuan2017}], would further suppress $\kappa_{l}$ from 1.7~W/mK to 1.5~W/mK, which brings the predicted value to an excellent agreement with experimentally measured $\kappa_{l}$ of 1.4~W/mK in Ref.~[\onlinecite{LiJuan2017}].

In order to assess the potential to enhance $zT$ by reducing $\kappa_{l}$ through nanostructuring, we examine the mean free path (MFP$\equiv\Lambda$) cumulative lattice thermal conductivity
\begin{equation}\label{eq:kappamfp}
\kappa_{l}^{\alpha\beta}(\Lambda)= \frac{1}{NV}\sum_{\lambda}^{|v_{\lambda}^{\alpha}|\tau_{\lambda} < \Lambda} C_{\lambda} v_{\lambda}^{\alpha} v_{\lambda}^{\beta} 	\tau_{\lambda},
\end{equation}
as shown in Fig.~2(e). We see that (1) the maximum MFP is about 100 nm, whether or not four-phonon scattering is included, (2) major heat-carrying phonon modes have MFPs in the range of 1 to 10 nm, and (3) four-phonon scattering significantly suppresses heat conduction from phonon modes with MFPs larger than 5 nm. To quantify the effect of nanostructure or grain size on $\kappa_{l}$, we additionally consider phonon-boundary scattering, the relaxation time of which is expected to be predominantly diffuse at 800~K,\cite{McGaughey2012} that is, $\tau_{\lambda, \text{Boundary}} = L/2v_{\lambda}^{\alpha}$, where $L$ denotes the sample size. Fig.~2(f) shows that sample size of about 10 nm is required to suppress $\kappa_{l}$ to about 1.0 W/mK when both three- and four-phonon scatterings are included.

Another widely-adopted strategy to reduce $\kappa_{l}$ is through alloying. Since experiments have reported significant improvement of thermoelectric performance of GeTe via alloying with cations such as Sb, Bi and Pb,\cite{WuDi2014,Li:2017aa,MinHong2018,Suresh2017,LiJuan2017,Suresh2016} we investigate their role in reducing $\kappa_{l}$. We treat the additional effects of alloying on the phonon spectra and $\kappa_{l}$ approximately, by accounting for mass disorder-induced extrinsic phonon scattering using Tamura's theory.\cite{Tamura1983} From Table~\ref{table:KappaGeTe}, we see that alloying GeTe with 10\% of either Bi or Pb reduces $\kappa_{l}$ to about 0.8 W/mK due to the large mass contrast compared with Ge, while alloying with 10\% Sb only slightly reduces $\kappa_{l}$ to 1.4 W/mK. 
Our results on Bi- and Pb-doped GeTe are overall consistent with experimental measurements\cite{Li:2017aa,Suresh2017,Suresh2016,WuDi2014} (see Table~\ref{table:KappaGeTe}) and the larger $\kappa_{l}$ observed in experiments may be due to the solubilities of Bi and Pb less than 10\%.\cite{WuDi2014,Suresh2016} However, our results for Sb-doped GeTe are significantly larger compared to experimental values.\cite{MinHong2018,LiJuan2017} It should be noted that there is also large discrepancy among different experiments (see Table\ \ref{table:KappaGeTe}). Since Sb has a solubility of about 10\% in GeTe and less lattice mismatch compared to Bi/Pb, other phonon scattering mechanisms beyond mass disorder scattering are required to explain the significantly reduced $\kappa_{l}$ in Ge$_{0.9}$Sb$_{0.1}$Te. As suggest by Perumal \textit{et al.},\cite{Suresh2017} other factors may include various nano/mesoscale structures such as defect layers, solid solution point defects, and nanoprecipitates. Moreover, there may be interplay between (Ge) vacancy concentrations and alloying elements, as recently identified in InSb,\cite{MAO2018189} which would also influence the value of $\kappa_{l}$. These factors may be explored in further investigations. 

\begin{table}[ht]
\centering
\caption{Computed $\kappa_{l}$ (W/mK) of pristine GeTe and Sb/Bi/Pb-alloyed GeTe at 800 K compared with experimental measurements at or near 800 K above $T_{c}$ (770~K in Ref.[\onlinecite{WuDi2014}], 710~K in Ref.[\onlinecite{Suresh2016}], 800~K in Ref.[\onlinecite{Li:2017aa}] and Ref.[\onlinecite{LiJuan2017}], 720~K in Ref.[\onlinecite{Suresh2017}] and  780~K in Ref.[\onlinecite{MinHong2018}]). The fraction of Sb/Bi/Pb is shown in parentheses. } 
\centering 
\begin{ruledtabular}
\begin{tabular}{p{1.0cm}  c  c  c  c} 
                & GeTe     & Ge$_{1-x}$Sb$_{x}$Te & Ge$_{1-x}$Bi$_{x}$Te & Ge$_{1-x}$Pb$_{x}$Te \\ 
\hline 
Exp.         & 1.36\cite{WuDi2014} & 0.72 (0.1)\cite{MinHong2018} &  1.08 (0.065)\cite{Li:2017aa} & 1.06 (0.13)\cite{WuDi2014} \\ 
                 & 1.44\cite{LiJuan2017}    & 0.76 (0.1)\cite{LiJuan2017}     &  1.28 (0.05)\cite{Suresh2017}   & \\ 
                 & 1.40\cite{Suresh2017}   & 1.15 (0.1)\cite{Suresh2017}     &  1.10 (0.1)\cite{Suresh2016}  & \\ 
\hline 
This study        & 1.70/1.50$^{a}$ & 1.42 (0.1) & 0.81 (0.1) & 0.82 (0.1) \\ 
\end{tabular}
\end{ruledtabular}
\footnotetext{With additional Ge vacancies at 2.1 atomic \% as in Ref.[\onlinecite{LiJuan2017}]}
\label{table:KappaGeTe} 
\end{table}


In summary, we have used a first-principles-based scheme that is capable of resolving lattice instability at finite temperature to model phonon properties of rocksalt $\beta$-GeTe. This scheme, combined with three- and four-phonon scatterings and Peierls-Boltzmann transport equation, enables us to model lattice thermal transport in $\beta$-GeTe at 800 K. Lattice thermal conductivities were computed and analyzed. The results show that the current approach is able to explain the dynamical stability of $\beta$-GeTe at 800 K and the experimentally-observed lattice thermal conductivities of both pristine GeTe and Bi/Pb-alloyed GeTe, thus allowing further improvement of $zT$ through phonon engineering.

\noindent\textbf{Supplementary Materials}

See supplementary material for detailed descriptions of the theory, computational parameters, interatomic force constants and various convergence tests.

\noindent\textbf{Acknowledgements}

This material is based upon work supported by Laboratory Directed Research and Development (LDRD) funding from Argonne National Laboratory, provided by the Director, Office of Science, of the U.S. Department of Energy under Contract No. DE-AC02-06CH11357. Use of the Center for Nanoscale Materials, an Office of Science user facility, was supported by the U.S. Department of Energy, Office of Science, Office of Basic Energy Sciences, under Contract No. DE-AC02-06CH11357. This research used resources of the National Energy Research Scientific Computing Center, a DOE Office of Science User Facility supported by the Office of Science of the U.S. Department of Energy under Contract No. DE-AC02-05CH11231.

\bibliography{GeTe}

\end{document}